\documentclass[useAMS,usenatbib]{mn2e}
\usepackage{epsfig}
\usepackage{psfig}
\usepackage{amsmath}
\usepackage{natbib}
\usepackage{graphicx}

\def\simle{\mathrel{
   \rlap{\raise 0.511ex \hbox{$<$}}{\lower 0.511ex \hbox{$\sim$}}}}

\newcommand{\be}{\begin{equation}}
\newcommand{\ee}{\end{equation}}
\newcommand{\bea}{\begin{eqnarray}}
\newcommand{\eea}{\end{eqnarray}}

\title[The Axis of Evil revisited]{The Axis of Evil revisited}
\author[Kate Land \& Jo\~{a}o Magueijo]
{Kate Land$^{1}$ and Jo\~{a}o Magueijo$^{2,3,4}$\\
$^1$ Astrophysics, University of Oxford, Denys Wilkinson Building,
Keble Road, Oxford OX1 3RH, UK\\
$^2$ Perimeter Institute for Theoretical Physics, 31
Caroline St N, Waterloo N2L 2Y5, Canada\\
$^3$ Canadian Institute for Theoretical Astrophysics, 60 St George
St, Toronto M5S 3H8, Canada\\
$^4$ Theoretical Physics Group, Imperial College, Prince Consort
Road, London SW7 2BZ, UK\\
Email contact: krl@astro.ox.ac.uk, jmagueijo@perimeterinstitute.ca}

\begin{document}

\date{Accepted xxx. Received xxx; in original form xxx}

\pagerange{\pageref{firstpage}--\pageref{lastpage}} \pubyear{2006}

\maketitle

\label{firstpage}

\begin{abstract}
In light of the three-year data release from
WMAP we re-examine the evidence for the ``Axis of Evil'' (AOE). We discover
that previous statistics are not robust with respect to the
data-sets available and different treatments of the galactic plane.
We identify the cause of the instability and implement an
 alternative ``model selection'' approach. A comparison to Gaussian isotropic 
simulations find the features significant at the 94-98\% level, depending on the 
particular AOE model. The Bayesian evidence finds lower significance, ranging 
from ``substantial'' at $\Delta(\ln E)\sim 1.4$, to no evidence for the most 
general AOE model.

\end{abstract}

\begin{keywords}cosmic microwave background\end{keywords}

%--------------------------------------------------------------

\section{Introduction}

The Wilkinson Microwave Anisotropy Probe (WMAP) has produced
spectacular high resolution all-sky observations of the Cosmic
Microwave Background (CMB), which have bolstered the case for the
$\Lambda$CDM concordance cosmological model~\citep{WMAP03cosmo,WMAP06cosmo}.
After the release of the first-year results~\citep{wmap03} there
was a flurry of studies into the Gaussianity and statistical
isotropy of the data, as these are fundamental predictions of
inflation theories. Reports of something awry have 
been obtained using a variety of 
techniques ({\it e.g.},~\cite{park,erik1,hbg,dondon,uscubic,hansen,erik2,vielva}). 
In this paper we focus on anomalies in the largest scale 
modes, after it was first noted that
the quadrupole ($\ell=2$) and octopole ($\ell=3$) appeared to be
correlated~\citep{oliveira}, and their power is suspiciously low. Much
work has focussed on the alignment and ``planarity'' of these two 
multipoles~\citep{copi05,schwarz,virgo}; but 
in~\cite{usevil} it was seen that the alignment actually extends
to the \emph{four} multipoles $\ell=2-5$, along the axis
$(b,l)\approx(60,-100)$. This feature has been dubbed the ``axis
of evil'' (AOE).

To be more precise the AOE expression has come to signify various
different things. Generally it is intended to denote any form of
statistical anisotropy, {\it i.e.}, a feature in the CMB
fluctuations which picks a preferred direction.  This can be
realized in many ways {\it e.g.}, multipole planarity (the
dominance of $m=\pm\ell$ modes along the preferred axis), or a
more general form of $m$-preference. In this respect it must be said
that while everyone agrees on the
presence of the ``axis of evil'' \emph{in the data}, its extent is
still debated. The expression is also sometimes
associated with the low power in the low $\ell$s.
This is quite inappropriate: while low power may be related to the
AOE (see~\cite{ustemplates}) there is nothing
``axial'' or anisotropic in a power spectrum anomaly per se.

There are two possible fault lines 
in the analysis leading to the
``axis of evil'' effect. The first concerns the integrity of the
data itself, {\it i.e.}, contamination from noise, systematics and
foregrounds. Comparison between the first-year (WMAP1) and third-year 
(WMAP3) data releases shows that the raw
data has hardly changed on large scales. However there are several
``all-sky'' renditions of the data and these do lead to
significant disparities: in this paper we show that this is true
regarding the intensity of the AOE, so that discussions should
emphasise not so much 1st V's 3rd year data, but the various
treatments of the galactic plane region.

The second fault line concerns the ``meaning'' of the detection,
and by this we 
%don't mean the recurring metaphysical brawls
%between Bayesianism and frequentism.  Rather we have in mind 
mean the robustness of the statistics used, and whether there is support
for planarity or more general $m$-preference. The frequentist 
formalism provides no clean way to penalise for extra
parameters or to weigh-up the detections against each other, or
the null hypothesis. Instead simulations are used to assess how
likely it is to get such a feature in a Gaussian statistcally isotropic 
(SI) CMB sky, 
but selection effects (by which we mean the tuning of the statistic or model 
to the data) are hard to account for. 
Here the confrontation of Bayesianism and frequentism becomes a very
practical matter.

We carry out this project as follows. 
In Section~\ref{freq} we re-examine the original frequentist AOE results for
various renditions of the WMAP1 and WMAP3 data, and we discuss
further the limitations of the original frequentist method, such
as its lack of robustness, at least with regards to $m$-preference 
AOE (as opposed to ``planarity''). In Section~\ref{bayes} we
follow a model comparison approach, and find that this is much more
robust when confronted with the different data-sets. 
Further we compare the evidence for the models; planarity and more 
complex $m$-preference. In Section~\ref{end} we summarise and discuss the 
results.

%---------------------------------------------------------------------------

\section{Instabilities of the frequentist statistics}\label{freq}

\begin{table*}\centering
\begin{tabular}{|l|cr|cr|cr|cr|c|}
\hline
 & $\ell=2$ && $\ell=3$ && $\ell=4$ && $\ell=5$ & & Mean\\
 \hline
 Map & $(b, l)$ & m & $(b, l)$ & m & $(b, l)$ & m & $(b, l)$ & m & inter-$\theta$
 \\\hline
%WMAP1 & (68.4, -82.7) & 2 & (13.5, 115.4) & 1 & (57.6, -161.6)  & 2& (47.7, -96.0)  & 3 & 53.5\\
LILC1 & (0.9, 156.7) & 0 & (63.0, -126.9) & 3 & (56.7, -163.7) & 2 & (48.6, -94.7) & 3 & 51.4 \\
TOH1 & (58.5, -102.9)& 2 & (62.1, -120.6) & 3 & (57.6, -163.3)  & 2& (48.6, -93.4) &  3 & 22.4\\
%TOHw1 & (56.7, -103.4) &  2  & (61.2, -120.6)  & 3 & (57.6, -163.3)  & 2 & (48.6, -93.4)  & 3 & 22.3\\
\hline%
TOH3 & (76.5, -134.0) & 2 & (27.0, 51.9) & 1 & (35.1, -130.6) & 1 & (47.7, -94.7) & 3 & 53.8 \\
%TOHw3 & (75.6, -133.2) & 2 & (27.0, 51.9) & 1 & (35.1, -130.6) & 1 & (47.7, -93.3) & 3 & 53.9 \\
WMAP3 & (2.7, -26.5) & 0 & (62.1, -122.6) & 3 & (34.2, -131.2) & 1 & (47.7, -96.0) & 3 & 53.7\\
%WMAP3bc & (70.2, -127.1) & 2 & (62.1, -122.6) & 3 & (34.2, -131.2) & 1 & (47.7, -96.0) & 3 & 25.0 \\
\hline
\end{tabular}
\caption{The ${\bf n}_\ell$ axes, in galactic coordinates $(b, l)$, and $m$ 
that maximise (\ref{aoestat}) for the multipoles $\ell=2-5$, 
for various all-sky renditions of
the first and third-year WMAP data. Note the low mean inter-angle
values for the TOH1 map, which indicate a strong
correlation between the multipoles ({it i.e.}, AOE). The dicontinuous nature
of the statistic causes the results to vary widely.}\label{aoetab}
\end{table*}

To assign an axis to each multipole, ~\cite{oliveira} proposed the following 
statistic: 
\be \label{qell} q_\ell=\max_{\bf n}\left[ \sum_m
m^2 |a_{\ell m}({\bf n})|^2\right], \ee where the $a_{\ell m}$s are
computed in the frame with ${\bf z}$-axis in direction ${\bf
n}$. This selects the frame dominated by the planar $m=\pm\ell$
modes.

In~\cite{usevil} we generalized this statistic to allow for any 
$m$ domination, {\it
i.e.}, not restricting ourselves to planar configurations, with the statistic: 
\be
r_\ell=\max_{m{\bf n}} \left[\frac{C_{\ell m}({\bf
n})}{\sum|a_{\ell m'}|^2} \right],\label{aoestat}\ee where $C_{\ell
0}({\bf n})=|a_{\ell 0}|^2$, and $C_{\ell m}({\bf n})=2|a_{\ell
m}|^2$ for $m>0$ (notice that 2 modes contribute for $m\neq 0$),
for the $a_{\ell m}$s computed in the frame with ${\bf z}={\bf n}$.
This produces three important quantities for each multipole: the
direction ${\bf n}_\ell$, the ``shape'' $m_\ell$, and the ratio
$r_\ell$ of the multipole's power absorbed by the mode $m_\ell$ in
the direction ${\bf n}_\ell$.

We extend the work of~\cite{usevil} by applying this statistic to
the following data-sets:
 
\noindent $\bullet$ The WMAP mission~\citep{wmap03} produced full sky CMB maps
from ten differencing assemblies (DAs).  They
also produced an ``internal linear combination'' (ILC) map. This 
assumes no external information about the foregrounds and combines 
smoothed frequency maps with weights 
chosen to minimize the rms fluctuations, using separate sets of 
weights for 12 disjoint sky regions. In the first-year data release the
WMAP collaboration advised that the ILC map be used only as a
visual tool. However, for the third-year
release a thorough error analysis of the ILC map was performed,
and a bias correction implemented~\citep{hinshaw06}. The resulting third-year ILC
map (herein WMAP3) is expected to be clean enough on
scales $\ell\simle10$ to be used without a mask. WMAP data is available from
http://lambda.gsfc.nasa.gov.

\noindent $\bullet$ Third-party maps include those of~\cite{TOH}, who produced
their own ILC map. Like above, an
``internal'' method is employed assuming only a black-body spectrum
for the CMB, but now the weights depend on scale (in harmonic
space) as well as galactic latitude. This is advantageous because
different sources of contamination dominate at different scales -
foregrounds at large scales, and noise at smaller scales.
%These maps are seen to be cleaner than the WMAP ILC map. 
As well as the cleaned
map, a Wiener filtered map is produced that, through a comparison
with the WMAP best estimates of theoretical $C_\ell$, adjusts the
power of the map so to suppress noisy fluctuations. We use their
first (TOH1) and third-year (TOH3)
cleaned-maps, all available from
www.hep.upenn.edu/~max/wmap.html.

\noindent $\bullet$ In an anlaysis of the ILC map-making method,~\cite{erikmap}
proposed a faster algorithm for the computation of the weights,
that employs Lagrangian multipliers to linearize the problem.
Although this produces identical results to that of the WMAP team,
and is indeed the method employed by the WMAP collaboration for
their third-year map, the authors applied it to the first-year
data using slightly different regions, thus producing a slightly
different ILC map (herein LILC1), available at
http://lambda.gsfc.nasa.gov. 
%Simulations were also produced~\citep{eriksims}.

\begin{figure*}
\begin{minipage}{8.0cm}
  \centerline{\psfig{file=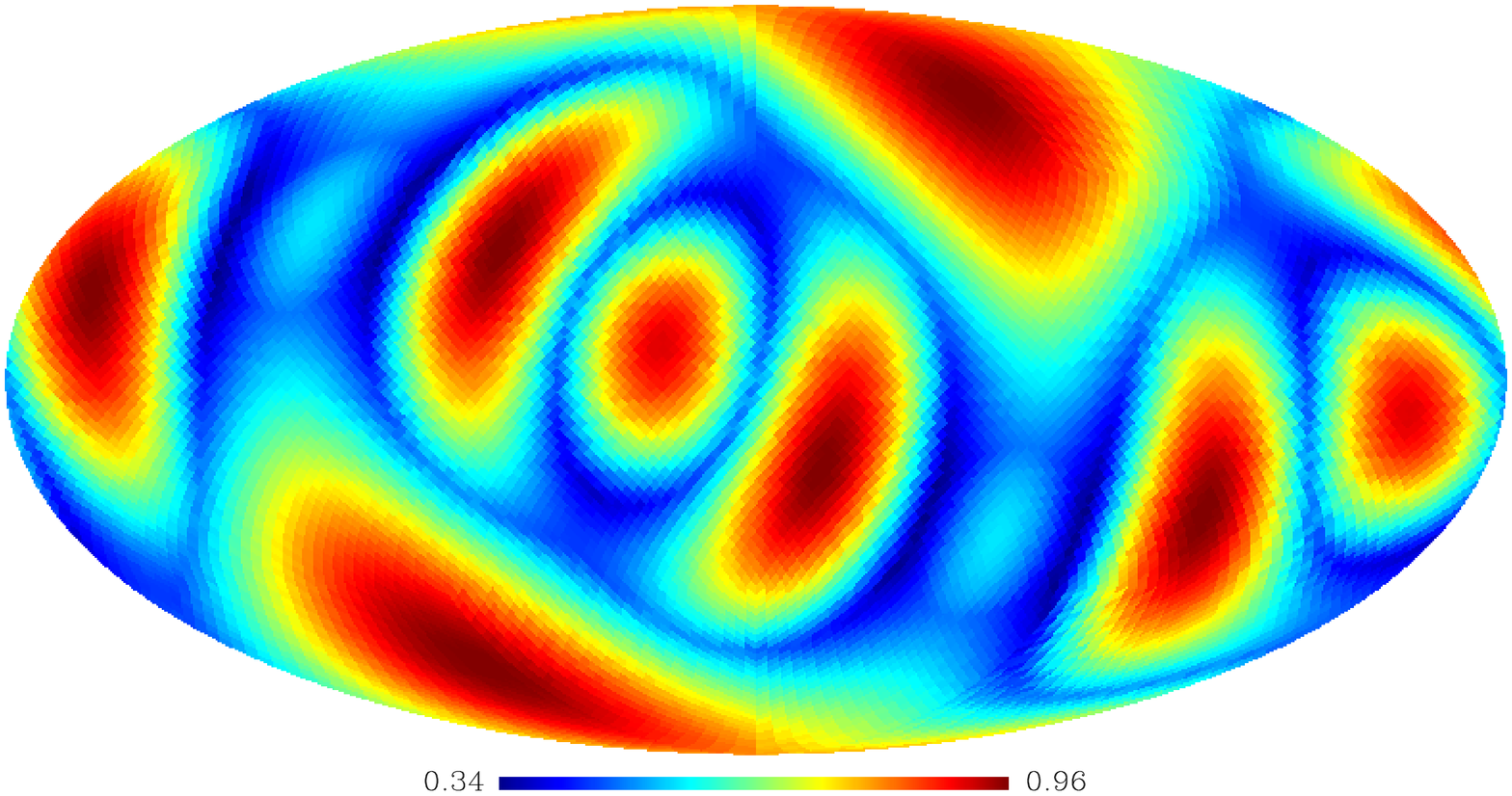,angle=90,width=7.5cm}}
  \centerline{\psfig{file=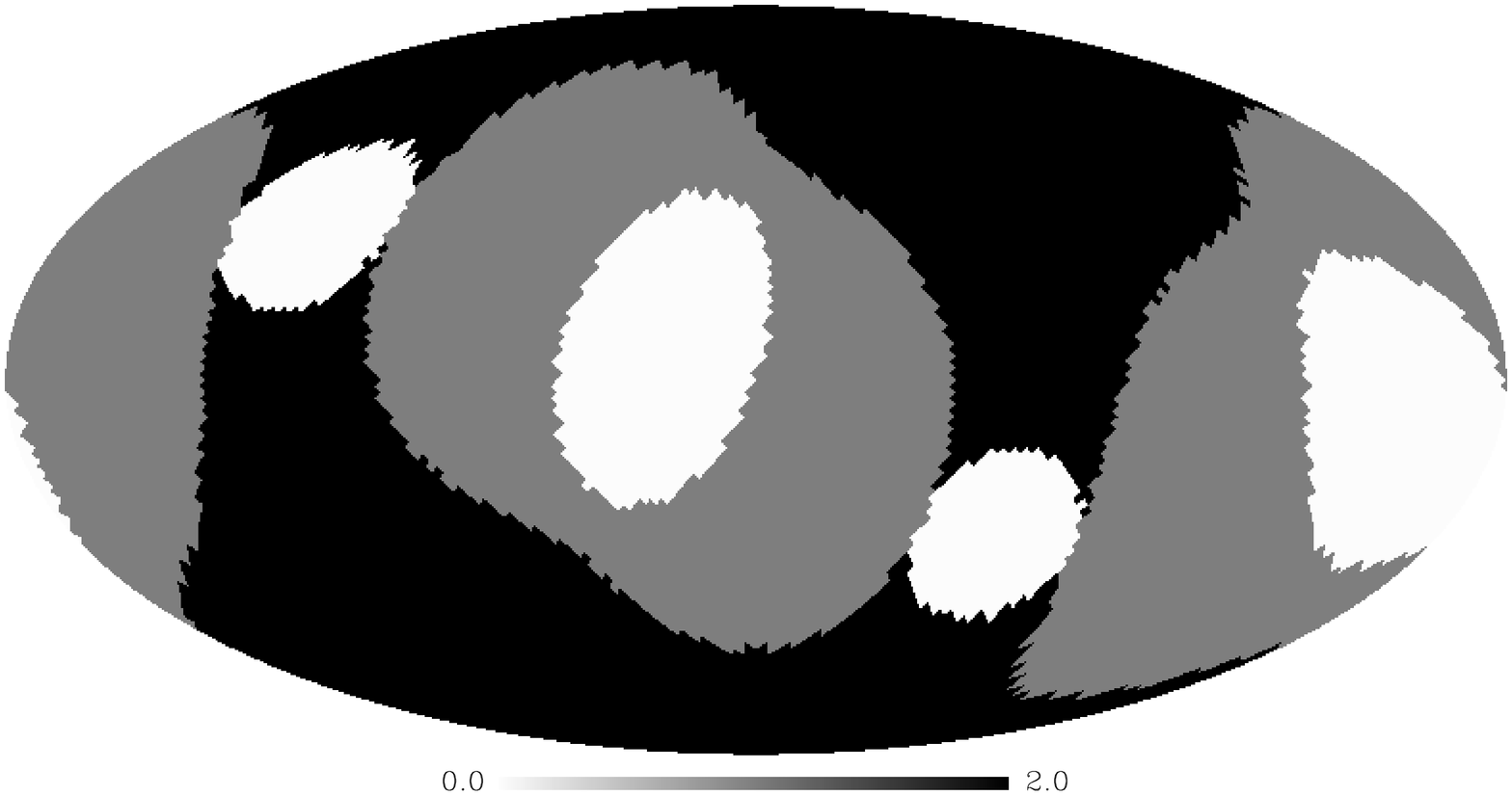,angle=90,width=7.5cm}}
\end{minipage}
\begin{minipage}{8.0cm}
  \centerline{\psfig{file=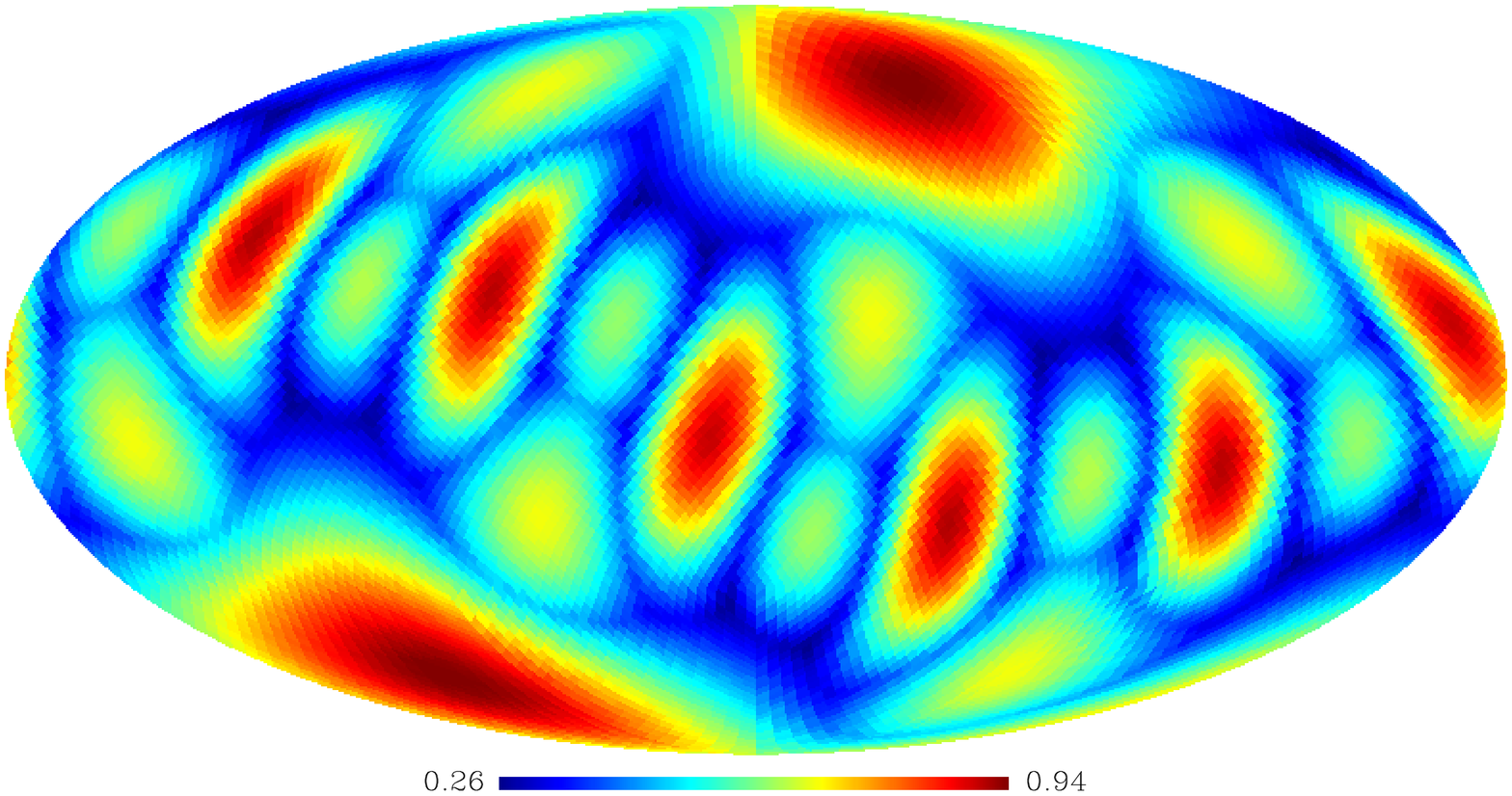,angle=90,width=7.5cm}}
  \centerline{\psfig{file=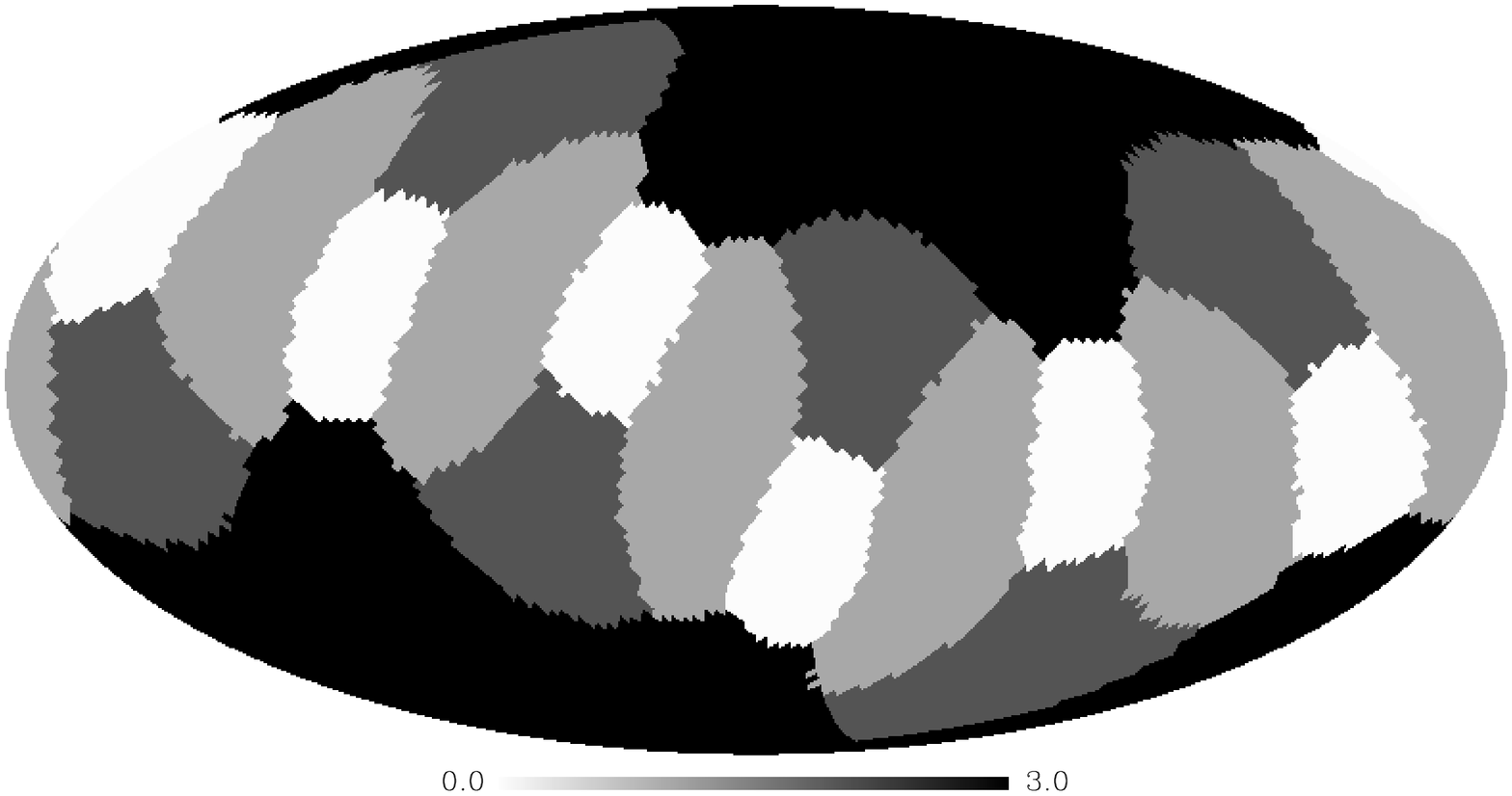,angle=90,width=7.5cm}}
\end{minipage}
\caption{The power ratio $R_\ell({\mathbf n})$ 
in the dominating $m$ mode
(above), and the $m$ value (below) for the quadrupole (left) and
octopole (right). The ``axis of evil'' statistic in 
(\ref{aoestat}) searches for the hottest spot in these
maps. We can see the close calls that cause the results to vary
widely in Table~\ref{aoetab}. Plotted in galatic coordinates and
Mollweide projection.}\label{aoefig}
\end{figure*}

There are of course the original frequency maps, which require a mask. 
However, for the task of assessing statistical isotropy
 we require full sky information, and thus
we only employ these ILC maps. 
%Although this process can never be perfect, in the future
%the community will require that these maps be clean enough for
%non-Gaussianity tests, as well as the extraction of the power
%spectrum. %Our analysis will touch on this issue as we explore the
%differences in these maps.

In Table~\ref{aoetab} we list the results obtained with frequentist 
AOE statistic (\ref{aoestat}) for the various
data-sets. It is clear that this statistic is not robust - very
similar maps can find very different results as indicated by the
final column. The expected inter-angle for isotropic axes is
1 radian ($\sim57^\circ$), thus a mean of $\sim 22^\circ$ is
remarkably low and a comparison to simulations puts this at the
99.9\% confidence level~\citep{usevil}. However, this result only
holds for two of the maps, and a small fluctuation in just one
multipole makes the ${\bf n}_\ell$ jump elsewhere. This highlights
one weakness of this statistic - its discontinuous nature.

In Fig.~\ref{aoefig} we visualize how ``close calls'' may arise,
explaining the discontinuities of the results in
Table~\ref{aoetab}. For the quadrupole and octopole of the TOH1
map, we plot the power ratio at the position ${\bf n}$
\be R_\ell({\mathbf n})=\max_m \frac{C_{\ell m}({\bf n})}
{\sum|a_{\ell m'}|^2}. \ee 
Thus the
``axis of evil'' statistic (\ref{aoestat}) picks out ${\bf n}_\ell$
as the position of the hottest spot from these maps (note
the degeneracy between $m=1,2$ for $\ell=2$ - we avoid this in
practice by taking just the $m=2$ solution). Below the $R_\ell$ maps
we plot the associated $m$ picked by $R_\ell$ for a given 
${\mathbf n}$.  We can now diagnose the instabilities in Table~\ref{aoetab},
by identifying close calls in the
competition for the hottest spot. For the quadrupole the $m=0$ and
$m=2$ modes, and for the octopole the $m=1$ and
$m=3$ modes are fighting a close battle. The overall mean
inter-angle (which measures the strength of the AOE) depends
closely on this battle, and thus the instability of this statistic. 
We should stress that the instabilities identified here do not seem to plague 
statistics for planarity~\citep{raf}.

%----------------------------------------------------------------------

\section{Model Comparison}\label{bayes}

The instabilities discovered appear to be cured by a model comparison treatment,
which allows for an evaluation of the evidence for $m$-preference in $\ell=2-5$, over simple
planarity for $\ell=2,3$.
Rather than computing a statistic from the maps 
({\it e.g.}, the mean inter-angle between the ${\mathbf n}$ for the various $\ell$),
the idea is to assess the ``evidence''  for a model encoding $m$-preference
or planarity, compared to the base model of statistical isotropy. 
We first outline the general formalism.
%This approach should be more robust as we can
%penalize for extra parameters and thus penalize for the jumping.
%We proceed with this in Section~\ref{bayes}, and first we outline
%some notation.

Let ${\cal L}$ be the likelihood of the data given a model, 
and $k$ the number of parameters of this model. The parameters 
should be tuned so to maximize the likelihood, or equivalently, 
to minimize the information in the data given the theory
(defined as $I(D|T)=-\ln({\cal L})$). However the real evidence 
should refer to the information in the data \emph{and} the theory: 
$I(D\cap T)=I(D|T) + I(T)$, where the 
information in the theory, $I(T)$, provides a penalization related
to the number of parameters. This matter is behind the 
``Occam's razor'' rationale~\citep{raf}, and the information 
criteria~\citep{lid}. 
%There is some controversy on the criterium for $I(T)$, the
%Aikaike information criterium (AIC) and the Bayesian information
%criterium (BIC) providing two formalisms that do not always agree.
According to the Aikaike informaiton criteria (AIC), 
the information in a theory is simply the
number of parameters, $k$. In fact, we will use a more accurate 
form, which is especially important for small sample size, 
$I^{AIC}=k+\frac{k(k+1)}{N-k-1}$, where $N$ is the
number of data points being fit~\citep{AICC}. 

An alternative approach to the problem of penalization is to compute 
the Bayesian evidence, 
\be E=\int {\cal L}(D|\theta,M)\; \Pi(\theta){\rm d}\theta=P(D|M),\ee
where $\Pi$ are the priors on the parameters $\theta$ for the model 
in question (see {\it e.g.},~\cite{trotta} for a review). Bayes theorem tells us how this is related to the probability of a model $P(M|D)$, and it provides an effective penalization by computing the \emph{average} 
of the likelihood over this expanded parameter space.
As an approximation to the logarithm of the 
Bayes factor, $B\equiv E_1/E_0$, we will compute the 
Bayesian information criteria (BIC), 
$I^{BIC}=\frac{k}{2}\ln N$ (confusingly this is not actually 
related to information-theoretic methods).  

%We see that this provides the proability of a model $P(M|D)\propto E $, 
%and thus the Bayes factor $(E_1/E_0)$ measures how much we prefer one model over another 
%(assuming uniform priors on the models).

The evidence $H$ for a theory $T_1$ is then defined as the 
decrease in the information of data and theory 
when it is compared with 
a null hypothesis $T_0$:
\bea H &=& I(D\cap T_0) - I(D\cap T_1) \nonumber\\
% &=& I(D|T_0) - I(D|T_1)  +  I(T_0)-I(T_1)\nonumber\\
 &\equiv& H_f           - H_p  ,\eea
where $H_f$ measures the improvement in the fit $H_f = \ln({\cal
L}_1)-\ln({\cal L}_0)$, and $H_p$ is the extra penalization we
have in our new theory.

In the language of the Jeffreys' scale~\citep{jeff,lidms}
 $\ln(B)$, or $H$, between 1 and 2.5 signals substantial evidence, 
between 2.5 and 5 signals strong evidence, and ``decisive'' evidence requires 
$\ln(B)>5$. However, for these rules of thumb to apply to the IC methods, 
various conditions should be met. For example the AIC assumes Gaussianity of the likelihood with respect to the parameters, while 
the BIC assumes independent identically distributed data points.
We will therefore compare these results to those from statistically isotropic 
Gaussian simulations, in Section~\ref{hybrid}. 
We will also compute the Bayes factor, for comparison with the 
BIC approximation, and the frequentist results.

\begin{table}\centering
\begin{tabular}{|l|lll|l|lll|}
\hline
Data-set& ($b$ & $\quad l$) & $\epsilon$  &$H_f$  &$H^{AIC}$ & $H^{BIC}$ & $\ln B$ \\
\hline
LILC1 & 63& -120 & .042 & 6.51  & 2.01 & 2.78 & 1.36\\
TOH1 & 61& -113 & .032 & 7.48  & 2.98 & 3.75 & 1.85 \\
%TOHw1& 60& -113 & .032 & 7.32  & 2.82 & 3.59 & 1.71 \\
%WMAP1&&&&&& 1.57\\
\hline
TOH3 & 74& -129 & .018 & 6.97  & 2.47 & 3.24 & 1.27 \\
%TOHw3& 73& -128 & .018 & 6.94  & 2.54 & 3.21 & 1.25 \\
WMAP3   &64 &-123 & .043 & 6.49   & 1.99  & 2.76 & 1.32 \\
%WMAP3bc   &65 &-123 & .038 & 6.89   & 2.39  & 3.15 & 1.53 \\
\hline
\end{tabular}
\caption{The maximum likelihood improvement, $H_f$, and best-fitting 
parameters for the planarity model ({\it i.e.}, 3 extra parameters), 
from various all-sky 
renditions of the WMAP data. We consider 
the evidence from AIC and BIC methods as well as the 
Bayes factor ($\ln B \equiv \Delta\ln E$).} \label{Htab1}
\end{table}

\subsection{Planarity model}
It was shown in~\cite{raf} that the planarity of the $\ell=2,3$
multipoles is supported by a Bayesian analysis. 
The model used to assess the evidence for planarity is
based on the diagonal covariance matrix: \be\label{planar}
{\langle |a_{\ell m}|^2 \rangle}({\bf n}) = c_\ell
\left(\delta_{\ell |m|}+ \epsilon (1-\delta_{\ell |m|})\right) \ee
where ${\bf n}$ and $\epsilon$ are the free parameters of the
model (in addition to $c_\ell$ that is common with the isotropic
model, but of a different value), with $\epsilon\le 1$. 
We use the same $\epsilon$ and ${\bf
n}$ for both multipoles, 
so that $k=3$, $N=12$, $H^{AIC}_p=4.5$, and
$H^{BIC}_p=3.73$. In Table~\ref{Htab1} we list
the parameter values that maximize the likelihood, together with
$H_f$ and $H$ following the AIC and BIC methods.

We also compute the Bayesian evidence
and record $\ln(B)$ in Table~\ref{Htab1}. We do this via brute force 
integration, and for the base model ($\left<|a_{\ell m}|^2\right>=c_\ell$) 
we use a uniform prior on $c_\ell$; $0\le c_\ell\:\ell(\ell+1)/(2\pi)\le 3000 \mu K^2$. 
For the planarity 
model we use uniform priors on $\epsilon \in [0, 1]$, and on 
$c_\ell$; $0\le c_\ell\:\ell(\ell+1)/(2\pi)\le 
(2\ell+1)\times3000 \mu K^2$, with the further constraint 
$\bar{c}_\ell\:\ell(\ell+1)/(2\pi)\le 3000 \mu K^2$ where $\bar{c}_\ell$ is the average 
$\bar{c}_\ell=\sum_m \left<|a_{\ell m}|^2\right>/(2\ell+1)$.

As before~\citep{raf} we find that variations between different
galactic plane treatments lead to only small variation in $H_f$. However, 
different evidence measures reach different conclusions. All the measures 
find at least substantial evidence for the planarity model, however the AIC 
and BIC appear to significantly overestimate this evidence compared to the $\ln(B)$ 
result. We refer to Section~\ref{hybrid} for a frequentist 
assessment of significance, through an analysis of $H_f$ from simulations.

\subsection{General $m$-preference model}
Using the same formalism we now revisit the debate on the extent
of the AOE, {\it i.e.}, $m$-preference as opposed to planarity. 
In the Bayesian formalism the matter can be addressed by replacing
the the covariance matrix (\ref{planar}) by
\be\label{model}%
{\langle |a_{\ell m}|^2 \rangle}({\bf n}) =c_\ell (\delta_{m'|m|}+
\epsilon (1-\delta_{m'|m|}))
\ee%
where ${\bf n}$, $\epsilon$ and $m'(\ell)$ are the free parameters of
the model, with $\epsilon\le1$. We find that \emph{if we analyze each $\ell$
separately} we rediscover the instabilities reported in
Section~\ref{freq}. In Table~\ref{Htab2} we take TOH1 for
definiteness, and present the winning $m'$, its associated $(b, l)$
and $H_f$; and also the runner up in cases where we get close
calls in maximising $H_f$. We see that the Bayesian analysis, in
this set up, merely confirms the $\ell=2$, $m'=0,2$ and the
$\ell=4$, $m'=0,2$ instabilities.

However, a totally new perspective into these instabilities now makes
itself known. $H_f$ only becomes the real evidence $H$ after it is
degraded by the penalization $H_p$, related to the number of
parameters of the model. If we allow each $\ell$ to choose its own
parameters then the overall $H_f$ is large (the sum total) but the
penalization is prohibitive as each multipole has 3 parameters.
%As explained in *** in the context of planarity,
%there is never an anomaly in $H$ (rather than $H_f$) if we choose different ${\bf n}$
%and $\epsilon$, $\ell$ by $\ell$. It is the fact that the chosen ${\bf n}$ are so similar
%that ``does the damage'', because we can choose a single, common ${\bf n}$
%for all the $\ell$s with minimal deterioration in $H_f$, but with striking savings
%in the penalization $H_p$.
Thus in optimizing $H$ we wish to reduce the number of parameters
by always seeking a common axis ${\bf n}$ for all $\ell$ in
(\ref{model}). This immediately removes the instabilities found in
the frequentist formalism, by effectively penalizing for jumping
between close calls, when one choice leads to a better {\it common}
set of parameters.

\begin{table}\centering
\begin{tabular}{|ll|lll|l|}
\hline
$\ell$ & $m'$ & ($b$ & $\quad l$) & $\epsilon$  &$H_f$  \\
\hline
2 & 0 &  6  & 157 & 0.027 & 3.47\\
2 & 2 &  59 & -103 & 0.030  & 3.09\\
\hline
3 & 3 &  62 & -120 & 0.025 &5.06\\
\hline
4 & 2 &  58 & -163 & 0.041&5.07\\
4 & 0 &  43 & -98  & 0.043&4.02 \\
\hline
5 & 3 &  49 &  -93& 0.026 &7.65\\
\hline
\end{tabular}
\caption{The maximum likelihood improvement, $H_f$, for a dominating 
$m$-mode model in the TOH1 
map, where each multipole can select its own axis, $\epsilon$, and
$m'$. Where there is a close call, the runner up $m'$ is also
listed.} \label{Htab2}
\end{table}

Take for example $\ell=2$. We have that
$m'=0, 2$ are close competitors in the optimization of $H_f$;
however only $m'=2$ picks an axis that is roughly aligned with the
preferred axis for the other multipoles. So only $m'=2$ permits a
large saving in $H_p$ ($H_p=2$ per axis, using, say, the AIC) with
only small deterioration in $H_f$. An instability would only arise
if $m'=0$ improved $H_f$ by an extra 2 when compared with $m'=2$.
The penalization forces the multipoles to chose common parameters,
at the risk of decreasing the fit a little.
Thus, in order to maximize $H$---and not only $H_f$---we should
select a common ${\bf n}$ for $\ell=2-5$, and the complete result
(for the same data-set) is presented in Table~\ref{Htab3}.

\begin{table*}
\begin{center}
\begin{minipage}{5.5cm}
\begin{tabular}{|ll|lll|l|}
\hline
$\ell$ & $m'$ & ($b$ & $\quad l$) & \quad$\epsilon$  &$H_f$  \\
\hline
2 & 2 & 49 & -96 & 0.052 & 2.33\\
3 & 3 & 49 & -96 & 0.108 & 2.01\\
4 & 0 & 49 & -96 & 0.058 & 3.21\\
5 & 3 & 49 & -96 & 0.028 & 7.34\\\hline
2-5 & --- & 49 & -96 & ------ & 14.89\\
\hline
\end{tabular}
\caption{The maximum likelihood improvement, 
$H_f$, for a dominating $m$-mode in the TOH1
map, where each multipole can select its own $\epsilon$ and $m'$,
but a common favoured axis is found.
%We also include the total $H_f$ obtained from combining all the multipoles.
} \label{Htab3}
\end{minipage}\hspace{0.8cm}
\begin{minipage}{11.0cm}
\begin{tabular}{|l|llll|l|ll|ll|}
\hline%
 Data-set& ($b$ & $\quad l$) & $\epsilon$  & $m$'s & $H_f$
&$H^{AIC}$ & $H^{BIC}$  & $\ln B$  & $\ln B_{23}$ \\%$H^{BIC}_{(+)}$\\
\hline
LILC1 & 48 & -100 & .077 & 2303 & 11.46 & 2.13 & -0.67 & -1.43 & -0.17  \\%1.34 \\ 2.68
TOH1 & 49 & -96 & .051 & 2303 &14.54  & 5.21 & 2.41 & 0.80  & 0.11   \\%4.42\\ 5.76
%TOHw1& 49 & -96 & .045 & 2303 & 15.25 & 5.92 & 3.12 & be  & 0.06   \\%5.13\\ 6.47
\hline
TOH3& 48 & -97 & .073 & 2303 & 11.57 & 2.24 &  -0.56 &   -1.15  & -0.21  \\%1.45\\ 2.79
%TOHw3& 48 & -97 & .067 & 2303 & 12.08 & 2.75 &  -0.05  & be  & -0.20  \\%1.96 \\ 3.30
WMAP3& 48 & -100 & .072 & 2303 & 12.10 & 2.77 &  -0.03 &  -1.01 & -0.18  \\%1.98 \\ 3.32
%WMAP3(bc)& 48 & -100 & .072 & 2303 & 12.01 & 2.68  & -0.12 & be & -0.08   \\%1.89 \\ 3.24
\hline
\end{tabular}
\caption{The maximum likelihood improvement, $H_f$, for the $m$-preference 
model with a common axis and common $\epsilon$
between the four multipoles $\ell=2-5$, and the variable $m'$, for
various data-sets ({\it i.e.}, 7 extra parameters). 
We consider the evidence $H$ using AIC,
BIC methods, as well as the Bayes factor ($\ln B$). We 
also compute the Bayes factor for just $\ell=2,3$.} \label{Htabm}
\end{minipage}
\end{center}
\end{table*}

In order to mimic the full treatment in~\cite{raf} we should also
seek a common $\epsilon$, thus reducing the number of parameters
further. This can be done via the method of
Lagrange multipliers, {\it i.e.}, by maximizing%
\bea H_{Tf}&=&\sum_{\ell,i}{\frac{N_{\ell
i}}{2}}{\left[{\frac{\sigma_{S\ell i}^2}{\sigma_{\ell i}^2}} +\ln
\sigma_{\ell i}^2\right]} -\lambda_1
[\sigma^2_{21}\sigma^2_{32}-\sigma^2_{31}\sigma^2_{22}]
\nonumber\\
&-&\lambda_2
[\sigma^2_{31}\sigma^2_{42}-\sigma^2_{41}\sigma^2_{32}] -\lambda_3
[\sigma^2_{41}\sigma^2_{52}-\sigma^2_{51}\sigma^2_{42}] \eea where
$i=1,2$ indexes the sub-samples for the $m$-modes with the
large and small variance respectively, with $N_{\ell i}$ modes
 and sample variance $\sigma_{S\ell
i}$. The solutions for the variance $\sigma_{\ell i}$ are constrained 
such that $\sigma_{\ell 2}/\sigma_{\ell 1}=\epsilon$, to fit with 
our model (\ref{model}). This has solution
\be%
\sigma^2_{\ell i}=\frac{\sigma_{S\ell i}^2}{1 -\frac{2
\alpha_{\ell
i}}{N_{\ell i}}} \nonumber \ee%
 with $\alpha_{2i}=\pm A$, $\alpha_{3i}=\pm
(-A+B)$, $\alpha_{4i}=\pm (-B+C)$ and $\alpha_{5i}=\mp C$, where
$A,B,C$ are solutions of the 3 quadratic equations expressing
$\epsilon_2=\epsilon_3=\epsilon_4=\epsilon_5$.

The results are presented in
Table~\ref{Htabm}. For all of the data-sets the choice $m'$s are the same
(as opposed to the frequentist statistics), and the preferred
common axis is remarkably robust. The common parameter $\epsilon$
and $H_f$ are also reasonably stable. Thus as far as choice of
statistics V's available data-sets are concerned we have found an
improved formalism and a robust set of best-fitting parameter values.

To compute the Bayes factor we use the same priors as before, with uniform 
priors on the additional $\{m'\}$ parameters, and we record 
$\ln(B)$ in Table~\ref{Htabm}.
 The AIC and BIC introduce penalizations of 9.33 and 12.13 respectively.
Regrettably at this point we see that the options for penalization spoil the
party, with the Bayes factor and BIC finding no evidence for the $m$-preference model (except for TOH1),
 while the AIC favors the $m$-preference model over the base model, and the planarity model 
 (except for TOH3).
We should perhaps not be overly disheartened by all this discord. It is
far from peculiar to the AOE effect: see for example the rather disparate
conclusions regarding evidence against scale-invariance ($n_S=1$) as
reported in~\cite{lid07}.
%In the following Section~\ref{hybrid} we examine the frequentist 
%approach further.

We note that the BIC gives us a simple tool to examine the effect of priors. 
If, for example, the model has a built in
positive mirror parity~\citep{deOliveira-Costa:1995td,Starobinsky:1993yx,usodd},
the number of possible $m'$ values is reduced, leading to a
lower penalization (10.12) for the same $H_f$ (only mirror positive modes are found 
in the data). This improvement of ~2 in the $H^{BIC}$ values will push the 
BIC (and probably the $\ln(B)$) result to favor this particular 
``positive reflection parity'' model 
over the base mode. But such a prior should be physically motivated.

%with the results ranging from substantial to decisive evidence!
%introducing some ambiguities. Undoubtedly using the AIC we find
%decisive evidence for $m$-preference AOE.
% Bruteforce use of the BIC leads to a very different
%picture, mainly because the penalization for the axis and
%$\epsilon$ jumps from $\frac{3}{2}\ln 12\approx 3.73$ for
%planarity to $\frac{3}{2}\ln 32\approx 5.20$.

%One may also question the way we treat the discrete variables
%$\{m'_\ell\}$. Reflecting on the number of computer bits taken to store it:
%$(\ln 2+\ln 2+\ln 3+\ln 3)$, makes the total penalization of $H_p^{BIC}=8.78$ used in 
%Table~\ref{Htabm}.
%If we make it simply the $\frac{1}{2}\ln$ of the number of
%data points used to determine each:
%%possible $m'$ values:
% $\frac{3}{2}\ln 32 + \frac{1}{2}(\ln 5 + \ln
%7 + \ln 9 + \ln 11)$, then this actually makes the penalization
%worse (9.27).

%---------------------------------------------------------------------

\subsection{Simulations}\label{hybrid}

\begin{figure*}
\centering
\begin{minipage}{5.5cm}
\centerline{\psfig{file=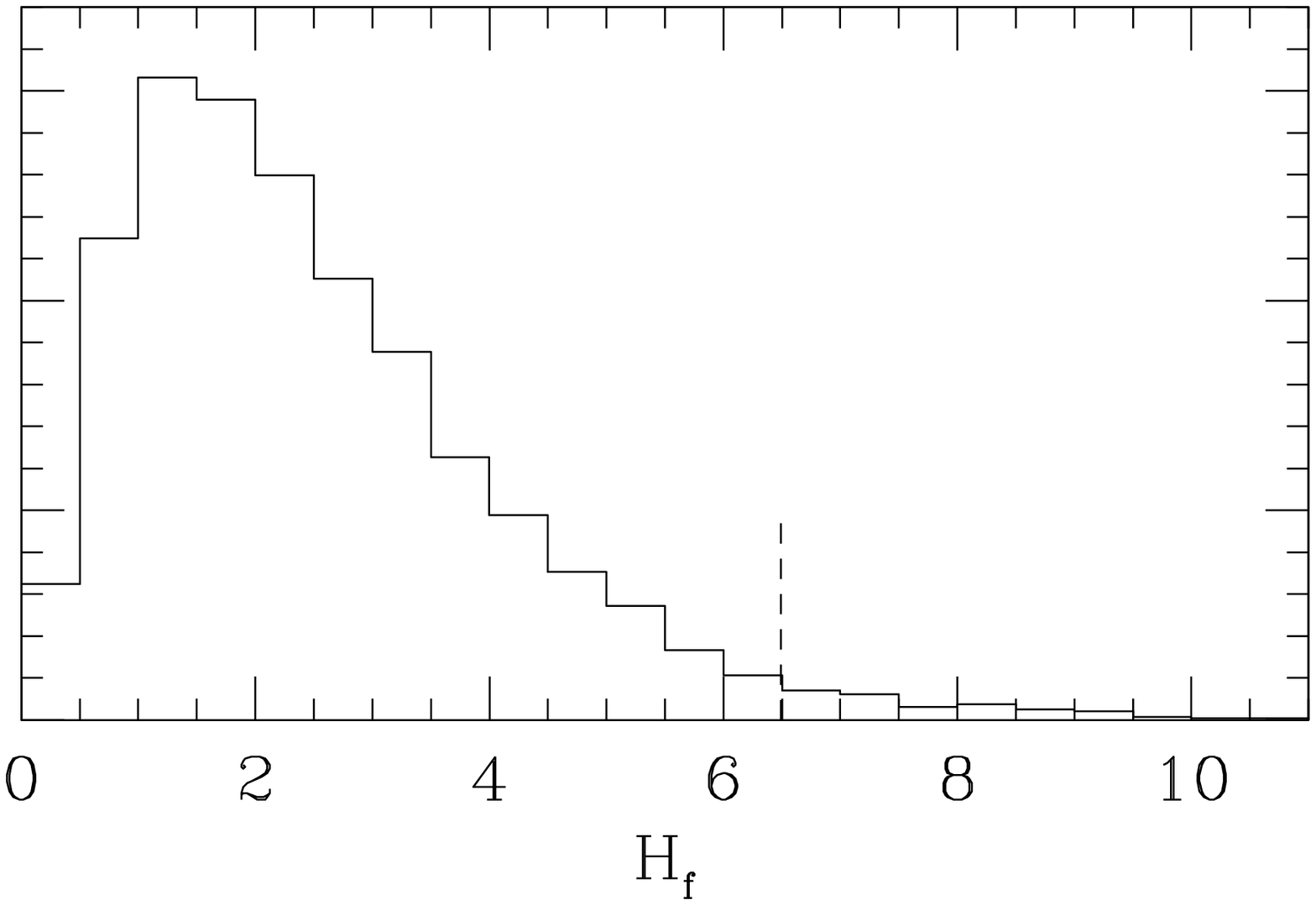,width=5.5cm}}\end{minipage}
\begin{minipage}{5.5cm}
 \centerline{\psfig{file=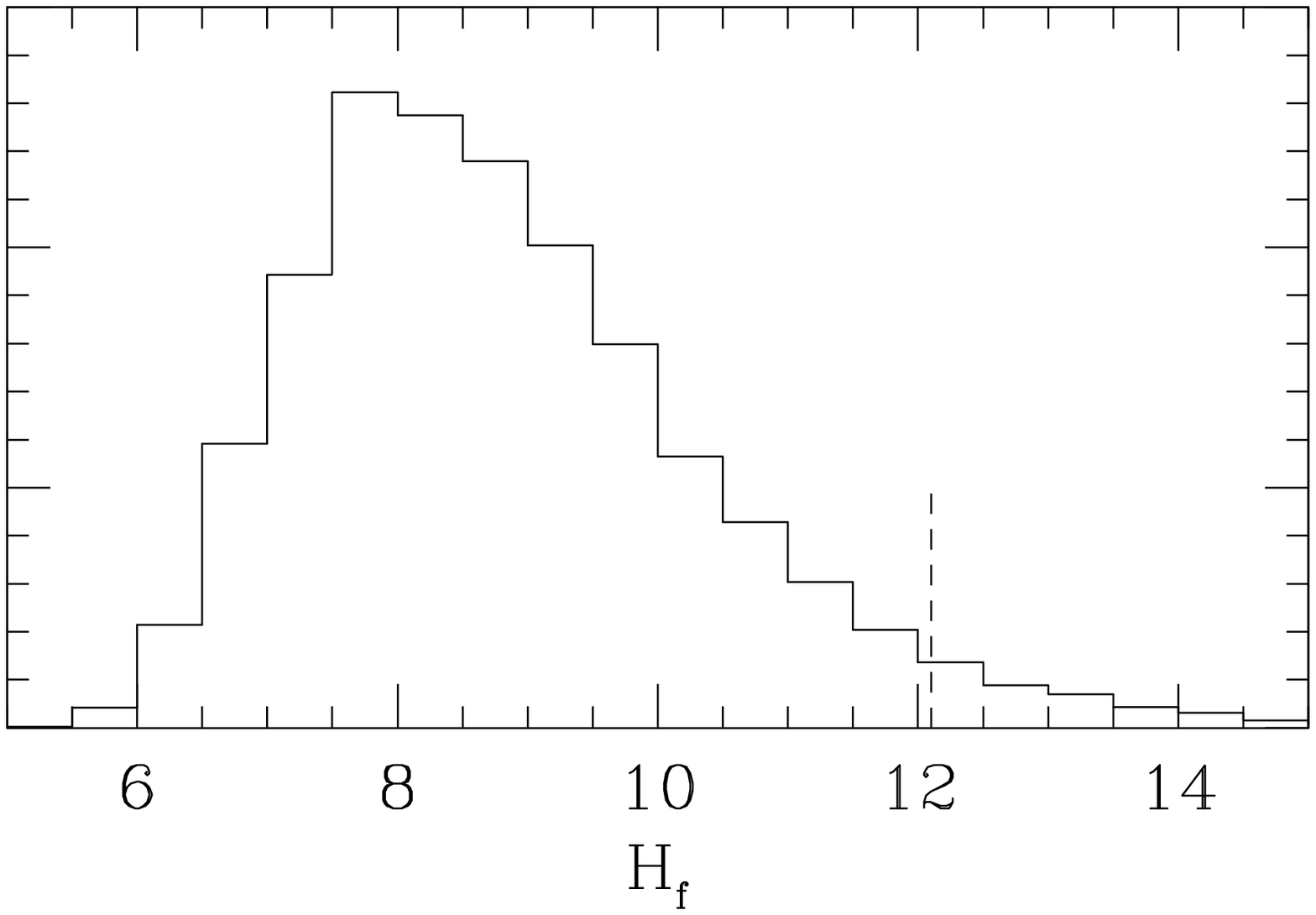,width=5.5cm}}
\end{minipage}\hfill
\begin{minipage}{6.5cm}
\centering
\begin{tabular}{llclc}
\hline
 Data-set & $H_f^P$ & \% & $H_f^m$ & \% \\
\hline
LILC1 & 6.51 & 2.69 & 11.46  &  6.90\\
TOH1  & 7.48 & 1.37 & 14.54  &  0.53\\
\hline
TOH3  & 6.97 & 2.02 & 11.57  &  6.35\\
WMAP3 & 6.49 & 2.71 & 12.10  &  4.21\\
\hline
\\
\\
\end{tabular}
\end{minipage}
\caption{The distribution of $H_f$ returned by 10,000 
Gaussian and isotropic simulations for the planarity model (left) and 
the general $m$-preference model (middle). We also plot the result 
obtained by the WMAP3 map (short dashed line). In the Table we list the 
percentage of simulations that find higher $H^x_f$ values for the planarity model (P) and the 
$m$-preference model ($m$). We stress that this approach does not take account of the relative 
complexities of the models.}
\label{hfplot}\end{figure*}

To assess (in a frequentist way) the significance of the maximum 
likelihood values, $H_f$, in 
Tables~\ref{Htab1} and~\ref{Htabm}
we compare our results to those from simulations. 
We stress that this is an 
alternative to the Bayesian method, for which the evidence is completely 
summarised by the Bayes factor, $\ln(B)$, 
with significance determined by the Jeffreys' scale
. The \emph{frequentist} approach to model selection in 
this case involves 
simulating data for the base model (Gaussian statistically isotropic (SI) 
CMB) and 
computing our $H_f$ ``goodness of fit'' statistic for the proposed models (Eqns (\ref{planar}) and (\ref{model})). We then obtain 
frequency plots for $H_f$ which indicate how well one would expect the 
proposed models to fit Gaussian SI CMB data. If the WMAP data 
finds a significantly better fit then we can conclude that the data 
is unlikely to be from a Gaussian SI model, at some confidence level (CL).

%Another approach, popular with some authors, consists
%of regarding the Bayesian goodness of fit $H_f$ (or its improvement
%when comparing two theories) as nothing but another frequentist statistic, 
%for which the significance of a measured value has to be assessed with 
%Monte Carlo simulations based on the null hypothesis (here a Gaussian
%isotropic process). While this makes perfect sense for a frequentist, 
%we stress that it is rather unorthodox from the point of view of the 
%Bayesian approach, for which $H_f$, combined with a penalization, 
%is the final word. We do not condone or condemn such a hybrid approach (or
%indeed any other philosophy in statistics), so we merely present the
%results of such an exercise in this Section. 

%Before proceeding, however, it is important to stress that one matter
%on which the hybrid approach certainly does not shed light is
%the issue of penalization, because the complexity of the theory
%does not enter in the proposed statistics at all. Should we reach a 
%discrepancy between the significance of an anomaly using the Bayesian
%or this frequentist approach that is not to be seen as yet another
%penalization rationale, but is merely a statement of a divergence
%between the frequentist and Bayesian result.

We use 10,000 Gaussian SI simulations, with the latest WMAP 
best fit $\Lambda$CDM power spectrum, to find the distribution of $H_f$ for the 
planarity model and the $m$-preference model. We plot histograms of the 
results in Fig.~\ref{hfplot}. This approach provides us with an alternative measure of 
the significance of our $H_f$ values, and in Fig.~\ref{hfplot} we list the percentage 
of the simulations that find a higher $H_f$ value. We see that the planarity model 
consistently finds significance at the 98\% level. The $m$-preference model 
generally has lower significance, at the 94-96\% level, except for the TOH1 map 
which finds very strong evidence for the $m$-preference model, at $>99$\% level. 
Note that it is this map that finds the $m$-preference AOE with the
original statistic (see Table~\ref{aoetab}).

These results are in agreement with the Bayesian approach ($\ln(B)$ and BIC), as the planarity model 
is favoured over the more general $m$-preference model except for TOH1. 
However, the Bayesian 
approach generally finds lower evidence for these models compared to the base model, and 
it actually finds no evidence for the $m$-preference model (except for TOH3). 
This reflects the well known fact that the Bayesian 
approach to model selection tends to set a higher threshold than frequentist approaches 
({\it e.g.},~\cite{trotta,muk2,linder}). Which result is more ``correct'' is a matter of personal opinion, however
 the more conservative Bayesian approach is often preferred in the field of cosmological model selection.

A disadvantage of the Bayesian approach is its 
sensitivity to priors, and its insensitivity to useless parameters 
that are unconstrained by the data. 
However, the frequentist approach can involve a large amount of 
computational time and can be prone to selection effects ({\it i.e.}, 
using a statistic pre-tuned by the data).
Consider that we could 
\emph{always} choose some convoluted complex statistic for which our data returns anomalously 
high (or low) values, compared to the simulations. Only the Bayesian approach can help 
here in imposing a suitable penalization, by averaging the likelihood over the 
extra parameter space. This ensures that a model is preferred only if the improvement in the fit 
merits opening up this extra dimension of parameter space.

The IC method provides another way of penalizing for the extra parameters, however we see that the AIC generally prefers the $m$-preference model (with the most parameters) to the planarity or base model - in disagreement with both the Bayesian and frequentist approach.

%The Bayesian approach 
%rightly accounts for this, and we see it tips the balance a 
%little towards the $m$-preference model.

%The reverse of the coin of ``selection'' is that one can always 
%criticize any frequentist statistics for simply not being
%the most appropriate for the detection of a given effect. A negative
%result is then a reflection on the poor quality of the measure used,
%rather than an actual feature of the data. One might criticize
%the confidence levels based on $H_f$ obtained in this Section on 
%these grounds, and favour the results obtained in the previous
%Sections. We leave it to the reader's inclinations to draw conclusions
%on these matters.

%We compare the approaches as much as one can. For the planarity model 
%the simulations find $<H_f>=2.54$, which is lower than the IC 
%penalization which is at least $3$.
%For the $m$-preference model the simulations find 
%$<H_f>=8.86$, which is in better agreement with the IC approach. 
%The tension between the approaches may reflect the small number of parameters 
%and data-points we are fitting, however, we must 
%stress that in no way does the frequentist approach account for the level of 
%complexity in the underlying model.

\section{Conclusions}\label{end}

We have highlighted weaknesses with the original AOE statistic (\ref{aoestat}) that 
probed $m$-preference for $\ell=2-5$. %, but not those 
%targeting  the planarity of $\ell=2,3$.
These are primarily: 1) lack of robustness:
small changes in the data produce very different best-fitting parameter values, {\it i.e.},
the statistics are discontinuous; 2) variations with data-set: it is hard to 
connect varying results to
 imperfections in the data or the statistic; 3) the need for simulations to assess
significance: no way of penalizing for extra
parameters or comparing competing theories on an equal footing, {\it e.g.}, 
planarity V's general m-preference. 

We have found an improved formalism by employing a model selection approach,
which cures the instabilities by favouring common parameters between the multipoles. 
The original instabilities were due to the existence of multiple solutions 
for a given multipole. But bringing in a penalization related to the number of 
parameters of the model enforces ``Occams Razor'' and selects solutions 
where parameters are common between the multipoles. We now find the 
best-fitting parameter values are robust.

The model selection approach also allows assessment of the
relative Bayesian evidence ($\ln B$) for the planarity model
(correlation between $\ell=2,3$, $m'=\ell$ modes) and the 
$m$-preference model (a
correlation between $\ell=2-5$, $m'$ not restricted). This
extends the work of~\cite{raf} where the low-$\ell$ low-power
evidence was assessed, as well as planarity for some data-sets.

Using the Bayes factor, and the BIC approximation, we find that there is 
substantial evidence for the planarity model, but no evidence for the $m$-preference model.
We also take a frequentist approach to the problem, and compare the ``goodness of 
fit'' ($H_f$) to those from Gaussian SI simulations. In agreement with the 
Bayesian approach, we find stronger evidence for the planarity model ($\sim 98$\% CL), than 
for the $m$-preference model ($\sim 95$\% CL). These results are in contradiction 
with the AIC approach which finds evidence for both models, and generally 
stronger evidence for the $m$-preference model. We think this demonstrates a weakness 
of this crude statistic, that does not appear to penalize enough for 
extra parameters.

The $m$-preference model is a more general version of the planarity model. It is therefore not surprising that the evidence for the planarity model is higher, as the parameter space is smaller while still including the best fitting model ($m'=\ell$). Likewise, we could restrict the $m'$ parameters to positive mirror parity modes and find a higher Bayes factor. But without a theoretical motivation for restricting the $m'$ parameters to these values it could be argued that this approach involves tuning our model (or equivalently - the priors) to fit the data. Therefore, the lower significance ($\sim 95$\%) result for the $m$-preference model is our more conservative result for the significance of the AOE in the WMAP third-year data. Note that the Bayes factor finds no support for this model, in multipoles 
$\ell=2-5$, nor for just $\ell=2,3$ (see last column of Table~\ref{Htabm}).

The higher significance returned by the simulations, compared to the Bayes factor, highlights an important difference between the Bayesian and frequentist 
approaches to model comparison. For 
some confidence level, the $\ln(B)$ threshold and frequentist $H_f$ threshold 
can disagree, with the Bayesian approach tending to be the more conservative
 - a phenomenon not unheard of when discussing ``2-sigma'' results.

%A weakness of this work is the use of full-sky maps, for which we know the foreground removal 
%can never be perfect. An improved approach would use a maximum likelihood method 
%to assess the most likely underlying $a_{\ell m}$ of the map, without using the 
%data in the galactic plane. 

\section*{Acknowledgements} We thank the many people who pestered us
with the question ``is it still there?'', and provided useful
conversations, most notably Andrew Liddle, Andrew Jaffe, Ofer
Lahav, Peter Coles and Carlo Contaldi. We are also grateful to
the referee, Hans Kristian Eriksen, for having suggested the approach
in Sec.~\ref{hybrid}, and championing the merits of the Bayesian 
evidence.

Our calculations made use
of the HEALPix package~\citep{healp} and were performed on COSMOS,
the UK cosmology supercomputer facility. KRL is funded by the Glasstone fellowship.

%\bibliographystyle{mn2e}
%\bibliography{shortBIB}

\label{lastpage}

\end{document}